\documentclass[twocolumn,prb,color,superscriptaddress,psfig,showpacs,amsmath,amssymb]{revtex4}

\usepackage{epsf}

\usepackage{float}

\usepackage{natbib}
\setcitestyle{square,numbers}

\usepackage{tabularx} 
\usepackage{graphicx} 
\usepackage{epstopdf}
\usepackage{epsf}
\usepackage{dcolumn}
\usepackage{bm}% bold math
\usepackage{color, xcolor}

\usepackage{multirow}

\usepackage[normalem]{ulem}

\usepackage{color,soul}
\setul{0.5ex}{0.5ex}
\setulcolor{red}

\begin{document}

\title{Kitaev Exchange and Altermagnetism in the Double Perovskite Sr$_2$CoTeO$_6$: An Ab Initio Study}

\author{Alexey V. Ushakov}
\affiliation{Institute of Metal Physics, Ural Branch of the Russian Academy of Sciences, Ekaterinburg 620137, Russia}

\author{Sergey V. Streltsov}
\affiliation{Institute of Metal Physics, Ural Branch of the Russian Academy of Sciences, Ekaterinburg 620137, Russia}

\date{\today}

\begin{abstract}
The Co$^{2+}$-based double perovskite Sr$_2$CoTeO$_6$ attracts attention due to its nontrivial magnetic properties. \textit{Ab initio} density functional theory calculations reveal a substantial ferromagnetic Kitaev exchange $|K|\sim J/3$ between nearest neighbors that, combined with the face-centered cubic lattice, leads to pronounced exchange frustration and places Sr$_2$CoTeO$_6$ among Kitaev materials with moderate Kitaev interaction strength. In addition, Sr$_2$CoTeO$_6$ is found to be altermagnetic with a sizable non-relativistic spin splitting, which can give rise to a transverse piezomagnetic effect and a nontrivial magneto-optical response.
\end{abstract}

\maketitle

%%%%%%%%%%%%%%%%%%%%%%
\section{Introduction\label{sec1}}
The Kitaev (spin) model~\cite{Kitaev2006} has become one of the most extensively studied subjects in recent decades~\cite{Takagi2019,trebst2022}.  Initially, only the honeycomb lattice and heavy $4d$ and $5d$ transition metal compounds - such as Na$_2$IrO$_3$ and $\alpha$-RuCl$_3$, which possess large spin--orbit coupling and have the specific electron filling $t_{2g}^5$ - were considered~\cite{Jackeli2009,Winter2016}. First, it was soon recognized that none of these materials constitutes an ideal realization of the model, owing to significant non-Kitaev terms such as the isotropic Heisenberg exchange $J$ and the symmetric anisotropic exchange constant $\Gamma$, see e.g.~\cite{Winter2016}. Second, it has long been established that the spin--orbit-derived $j_{\mathrm{eff}}=1/2$ doublet exists not only in the aforementioned iridates and ruthenates but also in many Co$^{2+}$-based compounds (i.e., those with a $d^7$ electronic configuration)~\cite{Abragam}. 
Many of these materials have layered structures with Co$^{2+}$ ions forming a honeycomb lattice, and investigations of their magnetic properties from the perspective of the Kitaev model have therefore intensified~\cite{Liu2018,Sano2018,maksimov2022}.

Another route by which Kitaev physics has advanced involves different geometries. For example, breaking the bonds in the honeycomb lattice, thereby reducing the dimensionality to 1D or quasi-1D, naturally enhances quantum fluctuations and stabilizes disordered regimes~\cite{agrapidis2018,maksimov2024}. On the other hand, materials with a triangular lattice~\cite{paddison2017,bordelon2019,razpopov2023,ortiz2023} and even face-centered cubic (fcc) structures~\cite{cook2015,aczel2016,revelli2019a,khan2019} have also been explored.  In the latter case, the physical realization is mostly in double perovskites with the general formula A$_2$BB'O$_6$, where only one of the two B sites (i.e., either B or B') is occupied by transition-metal ions.  In this geometry, two intermediate ligands belonging to two different BO$_6$ octahedra provide the exchange pathway between magnetic centers. 

\begin{figure}[!htbp]
\centerline{\includegraphics[width=0.9\linewidth]{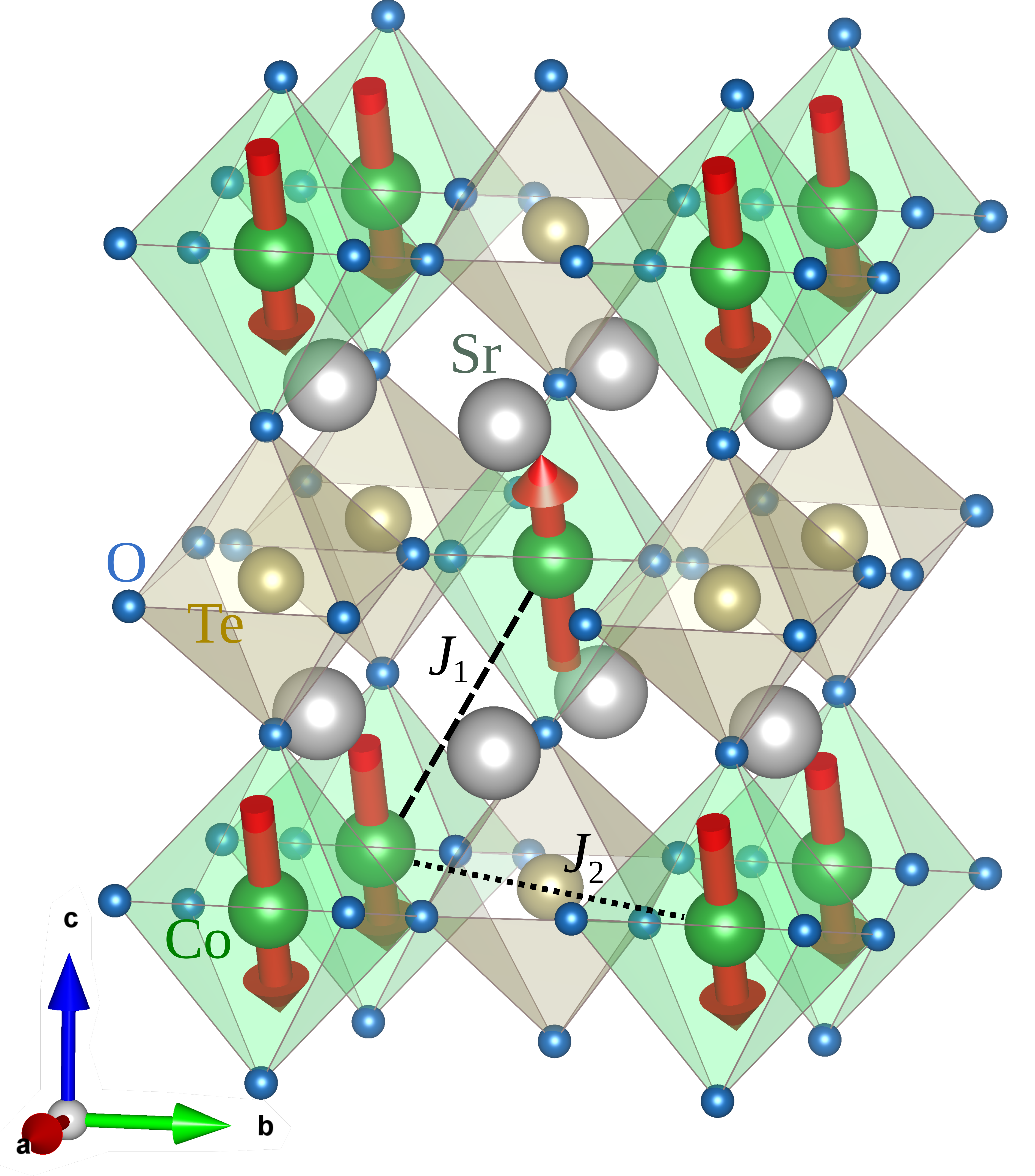}}
\caption{Experimental ($P2/n$) crystal and magnetic structure of Sr$_2$CoTeO$_6$. Nearest ($J_1$) and next-nearest ($J_2$) neighbor exchange bonds are shown by dashed and dotted lines, respectively. The figure was prepared using VESTA package~\cite{momma2011vesta}.\label{fig:str}}
\end{figure}

In the present paper we concentrate on a theoretical study of one of the fcc-lattice double perovskites, Sr$_2$CoTeO$_6$ with Co$^{2+}$. It orders magnetically below $T_N = 13$--$18$~K with type I antiferromagnetism (AFM): the ferromagnetic $ab$ planes stacked antiferromagnetically along the $c$-direction, see Fig.~\ref{fig:str}~\cite{augsburger2005,ortega-sanmartin2005,mathieu2011,tanaka2024}. The experimental local magnetic moments are consistent with the $2+$ oxidation state of Co being slightly suppressed to $2.2$-$2.4\mu_B$~\cite{augsburger2005,ortega-sanmartin2005}, which also agrees with density functional theory (DFT) calculations~\cite{ge2022}. They are tilted out of the $ab$ plane. An analysis of the specific heat is also consistent with Co$^{2+}$ and a $j_{\textrm {eff}} = 1/2$ doublet in the ground state~\cite{ortega-sanmartin2005}. The Curie-Weiss temperature is of the order of $-140$~K~\cite{augsburger2005,ortega-sanmartin2005}, which is much larger in absolute value than $T_N$, signalling substantial frustration. Recent measurement in large magnetic fields show an indistinct one-half magnetization plateau in fields $\sim$34-42 T~\cite{tanaka2024}.

At low temperatures, the crystal structure is characterized by the $P2/n$ space group. Interestingly, there are few structural transitions at high temperatures: at 373K (to $I2/m$), $\sim$650K (to $I4/m$) and 773K (to $Fm\bar3m$), but no symmetry modifications have been reported below $T_N$~\cite{zaraq2023}. Similar transitions have been observed in other double perovskites~\cite{Vasala2014a}. Optical measurements show that the material is an insulator with a band gap of $\sim$ 3.1-3.2 eV~\cite{zaraq2023,zaraq2026}.

In this work, we perform a theoretical study of the electronic and magnetic properties of Sr$_2$CoTeO$_6$ using {\it ab initio} calculations. The results confirm the experimentally established ground state of the system --- type-I antiferromagnetic ordering, where the spins lie in the $ac$ plane and are rotated from the $a$ axis by an angle of $\sim$58$^{\circ}$~\cite{ortega-sanmartin2005}. 
We also estimate the exchange integrals in the Heisenberg model using the total-energy and Green’s function methods, showing that the anisotropic contributions $K_1$ and $\Gamma_1$ are approximately 1/3 of the magnitude of the isotropic term $J_1$, which yield experimentally observed magnetic ground state of Sr$_2$CoTeO$_6$, while not being far from the spin-liquid regime
%We also estimate the exchange integrals in the Heisenberg model using the total-energy and Green's function methods, showing that the anisotropic contributions $K_1$ and $\Gamma_1$ are approximately $1/3$ of the magnitude of the isotropic term $J_1$, which may bring the system closer to the spin-liquid regime. 
The magnetic point group $2/m.1$ allows for $M$-type altermagnetism, as confirmed by calculations. Altermagnetism, in turn, induce a transverse piezomagnetic effect in the system and give rise to a spontaneous Hall effect with a nonzero $\sigma_{zx}$ component.

%%%%%%%%%%%%%%%%%%%%%%%%
\section{Calculation details}

%The crystal structure of CaCoSi$_2$O$_6$ at room temperature was taken from Ref.~[\onlinecite{Ghose-87}]. 
The band-structure calculations within the framework of DFT~\cite{kohn1964} were performed using the VASP package~\cite{kresse1996}. The generalized gradient approximation (GGA)~\cite{perdew1996} and the projector augmented wave (PAW) method~\cite{Bloehl1994} were applied. We utilized the exchange-correlation functional in the Perdew–Burke–Ernzerhof (PBE) form~\cite{Perdew1997}. The cutoff energy was set to 600 eV, and we used 4$\times$4$\times$8 {\it k}-point mesh for integration over the Brillouin zone according to the Monkhorst–Pack scheme~\cite{Monkhorst1976}.

In order to include correlation effects for the $3d$-electrons of the Co ions, we used the GGA+U calculation scheme as proposed in ~\cite{Liechtenstein1995}. The on-site Coulomb repulsion parameter $U$ was chosen to be 6 eV, and the Hund's rule coupling parameter $J_\textrm{H}$ was set to 0.9 eV~\cite{maksimov2022,maksimov2024}. The occupation numbers of the Co $3d$ states were obtained by integration within an atomic sphere with a radius of 1.302~\AA. For the estimation of the crystal field splitting and hopping integrals, a non-interacting GGA Hamiltonian for the Co $3d$ states was constructed via the Wannier projection procedure~\cite{korotin2008,korotin2015}.

Optical conductivity was calculated as described in Ref.~\cite{Gajdos2006}. Complex shift parameter was chosen to be $\eta$~=~0.1~eV. Higher lying states up to 30.3 eV above $E_\textrm{F}$ were included in these calculations.

\begin{figure}[htb]
\centerline{\includegraphics[width=\linewidth]{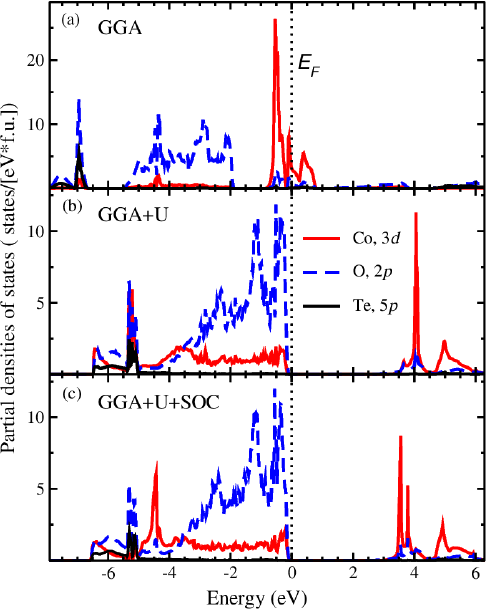}}
\caption{Partial densities of states of Sr$_2$CoTeO$_6$ obtained in nonmagnetic GGA, GGA+U, and GGA+U+SOC calculations. The Fermi energy is in zero..\label{fig:dos}}
\end{figure}

%%%%%%%%%%%%%%%%%%%%%%%%%%%%%%%%%%%%%
\section{Electronic and magnetic structure}
We start with an analysis of the simplest non-magnetic GGA calculations, with the results presented in Fig.~\ref{fig:dos}(a). One can see that the width of the Co $3d$ band is relatively small, $\sim$1.5 eV. This stems from the fact that the CoO$_6$ octahedra are not directly connected with each other.
The O $2p$ states form a broad band below the Fermi energy, spanning approximately $\sim$3.5 eV.
%and are strongly hybridized with the Co $3d$ states. 
These ligand bands are separated from the $3d$ band by an energy gap of about 1.2 eV. The Te $5p$ states are located about 7 eV below the Fermi level and are also separated from the O $2p$ states by an energy gap. 

The effective Hamiltonian constructed using the Wannier-function formalism for the Co $3d$ states allows one to estimate the crystal-field splitting between the $t_{2g}$ and $e_g$ states, $\Delta_\textrm{CFS} \sim 0.6$ eV. Structurally, the distortions in the CoO$_6$ octahedra are small, with the differences in Co--O bond lengths being less than 0.002~\AA. Thus, the intra-subshell splittings within the $t_\textrm{2g}$ and $e_\textrm{g}$ states are also small, amounting to about 0.03 eV.

Accounting for the on-site Coulomb repulsion on the Co ions makes the system insulating with an energy gap $E_\textrm{g} \sim 3.5$ eV, which is consistent with the experiment. Since the top of the valence band is formed by O $2p$ states and the bottom of the conduction band is dominated by Co $3d$ orbitals, this material can be classified as a charge-transfer insulator~\cite{ZSA}. 
%The Co $3d$ peaks become narrower, and the hybridization between Co $d$ and O $p$ states substantially increases, since occupied Co $3d$ orbital appears in the same region as ligand's $p$ states. 
The magnetic moment on Co is found to be 2.79~$\mu_B$, which is consistent with the high-spin Co$^{2+}$ state.

Spin-orbit coupling (SOC) is known to be important for Co$^{2+}$ ions having the $t_{2g}^5 e_g^2$ electronic configuration. In the case of an isolated ion, it splits the $t_{2g}$ levels 
into lower-lying doublet characterized by the effective total angular momentum $j_{\textrm {eff}}=1/2$ and higher-lying quartet $j_{\textrm {eff}}=3/2$~\cite{Abragam}. For our material, with factors such as crystal-field splitting, Coulomb interactions, and band effects, the situation is even more complicated. Nevertheless, as can be seen from Fig.~\ref{fig:dos}(c), SOC significantly affects the electronic structure modifying Co $3d$ density of states (DOS), see Fig.~\ref{fig:dos}(b,c). It also results in an orbital contribution to the magnetic moment, which amounts to $m_o =$~1.71~$\mu_B$, while the spin component is $m_s= $~2.77~$\mu_B$.

%The densities of states shown in Fig.~\ref{fig:dos}(b, c) correspond to the ground state of Sr$_2$CoTeO$_6$. 
Experimentally, this compound is a type-I antiferromagnet below the Néel temperature. as it has been mentioned above, the spin moments are parallel within the (001) plane, while adjacent planes are antiparallel in the type-I AFM ordering. In the type-II AFM ordering, the spins are parallel within the (111) plane, and adjacent planes are antiparallel.
The total energy of type-I AFM in GGA+U calculations with a cell containing 16 Co ions is lower by 7.4 meV/f.u. than that of the type-II structure. 

Moreover, experimental data indicate that the spin moments in Sr$_2$CoTeO$_6$ order in the $ac$ plane at an angle of $\sim 58^{\circ}$ relative to the $a$ axis. GGA+U+SOC calculations clearly show that for type-I ordering, the total energy of the structure with spins oriented along the [0.54, 0, 0.82] direction is lower than that of structures with spins pointing along the [100] and [001] by 1.9 and 1.0 meV/f.u., respectively. Thus, the GGA+U+SOC calculations not only reproduce the type-I magnetic structure but also confirm the experimentally determined orientation of the spin moments.

%%%%%%%%%%%%%%%%%%%%%%%
\section{Exchange interactions}

In double perovskites with the fcc lattice, each magnetic ion has 12 nearest neighbors and 6 next-nearest neighbors. In Sr$_2$CoTeO$_6$  the Co--Co bonds corresponding to the nearest neighbors are inequivalent: within the $ab$ plane there are two bonds of 5.59~\AA{} and two of 5.63~\AA{}, while there are also four distances of 5.605~\AA{} and four of 5.609~\AA{} connecting the perpendicular planes. 
%When calculating the exchange interactions, we cannot evaluate the $J$ integrals for each inequivalent Co--Co bond using the total-energy method; the computational cell would be too large (one would need 16 Co ions in the cell). Therefore, in our model we assume that the exchange interactions along all bonds between nearest-neighbor Co ions are equivalent (then one can restrict oneself to a cell containing 4 Co ions).

First, we estimated the isotropic exchange parameters in Sr$_2$CoTeO$_6$ within the classical Heisenberg model, using the Green's function formalism~\cite{korotin2008} and a Hamiltonian obtained by projection onto Wannier functions in the GGA+U approximation, which contains Co $3d$, O $2p$, and Te $5p$ states:
\begin{align}\label{eq:Heisenberg}
\begin{split}
H = \sum_{i>j} J_{ij} \mathbf{S}_i \cdot \mathbf{S}_j,
\end{split}
\end{align}
where $i,j$ index the lattice sites. The Green's function approach allows one to calculate the exchange integrals for each inequivalent Co--Co bond. Since our aim is to estimate the exchange parameters for the Kitaev-Heisenberg model, which operates with effective spins of 1/2, we use $S_{\textrm{eff}}=1/2$ for the Heisenberg model as well. 

The exchange integrals $J_1$ between the nearest-neighbor Co ions, see Fig.~\ref{fig:str}, vary from 92 to 108 K (AFM), depending on the Co--Co bond length, with an average $J_1$ of $\sim 100$ K (AFM). The Co--Co distances between the next-nearest neighbors are much longer (7.91--7.93~\AA{}) and $J_2$ is nearly the same for all six bonds, amounting to $J_2 \sim -4$ K (FM). The antiferromagnetic and ferromagnetic signs of $J_1$ and $J_2$, respectively, lead to type-I antiferromagnetic ordering~\cite{lefmann2001,diop2022}, which is consistent with experimental findings~\cite{augsburger2005,ortega-sanmartin2005,mathieu2011,tanaka2024}.
Using these exchange parameters, one can also calculate the Curie-Weiss temperature $\theta$, which is found to be $-188$ K. This is close to the experimental value of $-140$ K~\cite{ortega-sanmartin2005}, but slightly overestimated, as expected for the mean-field approximation.

In order to estimate the parameters of the anisotropic exchange interaction between nearest-neighbor Co ions, we used the Kitaev-Heisenberg spin model:
\begin{align}\label{eq:spin}
\begin{split}
H = &\sum_{\langle ij \rangle \in (\alpha, \beta, \gamma)}  J_{ij} \mathbf{S}_i \cdot \mathbf{S}_j + K_{ij} S_{i}^{(\gamma)} S_{j}^{(\gamma)} + {}\\
     &+ \Gamma_{ij} \bigl(S_{i}^{(\alpha)} S_{j}^{(\beta)} + S_{i}^{(\beta)} S_{j}^{(\alpha)}\bigr),
\end{split}
\end{align}
where $\alpha$, $\beta$, and $\gamma$ are the spin directions in the local coordinate system for each Co--Co pair. $J$, $K$, and $\Gamma$ are the isotropic Heisenberg, bond--dependent Kitaev, and bond--dependent off--diagonal terms, respectively.

In the double perovskite structure, one can set $\gamma = z$, $\alpha = x$, and $\beta = y$, where $(x, y, z)$ is the global coordinate system defined by the lattice vectors $\textbf{a}$, $\textbf{b}$, and $\textbf{c}$.
Then, using a unit cell containing 4 Co ions, one can consider three magnetic configurations ($\uparrow_\textrm{z}$, $\uparrow_\textrm{z}$, $\downarrow_\textrm{z}$, $\downarrow_\textrm{z}$), ($\uparrow_\textrm{z}$, $\downarrow_\textrm{z}$, $\uparrow_\textrm{z}$, $\downarrow_\textrm{z}$), and ($\uparrow_\textrm{z}$, $\downarrow_\textrm{z}$, $\downarrow_\textrm{z}$, $\uparrow_\textrm{z}$) -- in the GGA+U+SOC approximation. From the differences in their total energies, we obtain $J_1 = 70$~K (AFM) and $K_1 = -20$~K (FM) for $S_{\textrm{eff}}=1/2$, where the index $1$ refers to the nearest-neighbor exchanges.  It should be stressed that these are the estimates of the average exchange parameters, since the nearest-neighbor Co-Co bonds are slightly inequivalent, as discussed above. Decrease of the isotropic exchange $J_1$ with respect to what was obtained by GGA+U method is both due to modification of the electronic structure by the spin-orbit coupling and more accurate total energy, which includes all additional contributions to the total exchange interactions such as, e.g., via orthogonal ligand $p$ orbitals~\cite{streltsov2008a}. 

The $\Gamma_1$ parameter  cannot be calculated in this way, because in the fcc lattice with a unit cell containing four Co ions, the energy differences between any two magnetic configurations cancel the contribution of $\Gamma_1$. However, one can choose a single bond between nearest-neighbor Co ions in the cell and calculate $\Gamma_1$ using the four-energy method described in Ref.~\cite{Xiang-11}. The exchange integral $\Gamma_1$ is found to be 12 K. By including the isotropic next-nearest-neighbor exchange  $J_2$ calculated in GGA+U, the recalculated the Curie-Weiss temperature decreases to $-110$ K, which is consistent with the experiment. Mean-field estimation of the Neel temperature is 42K. Thus, one can conclude that  Sr$_2$CoTeO$_6$  exhibits an anisotropic contribution to the exchange interaction between nearest-neighbor Co ions that is approximately one-third of the magnitude of the isotropic term.

Interestingly, the calculated exchange parameters, $K/J_1=-0.29$ and $J_2/J_1 = -0.06$, place Sr$_2$CoTeO$_6$ rather close to the spin-liquid region; see the phase diagram of the Kitaev-Heisenberg model on the fcc lattice obtained in a pseudofermion functional renormalization group study, presented in Fig.~4 of Ref.~\cite{revelli2019a}. This explains the low N\'eel temperature and the large frustration parameter $f=|\theta|/T_N \sim 10$ observed experimentally. However, it should be noted that the exchange parameters strongly depend on the choice of the Hubbard $U$ and the intra-atomic Hund's coupling $J_H$~\cite{khomskii2014transition,khomskii2024}. While our selected values close to those used in the literature provide reasonable agreement with the experiment in terms of the Curie-Weiss temperature, further theoretical and experimental studies, including inelastic neutron scattering, are needed to construct a reliable spin model for Sr$_2$CoTeO$_6$.

Next, to determine the ground state of Sr$_2$CoTeO$_6$, we apply the generalized Luttinger--Tisza (LT) method implemented in the \texttt{Sunny.jl} package~\cite{Dah-25}. Using the exchange constants \(J_1\), \(K_1\), \(\Gamma_1\), and \(J_2\), we constructed the exchange matrices \(\hat{J}_{ij}\) in the global Cartesian frame, which coincides with the local axes of the CoO$_6$ octahedra. For nearest-neighbor bonds, the matrix is built according to the anisotropic exchange model
\begin{align}\label{eq:3}
\begin{split}
    \hat{J}_{ij} = J_1 \mathbb{I} + K_1\,\boldsymbol{\eta}\boldsymbol{\eta}^{T} + \Gamma_1\,(\boldsymbol{\xi}_1\boldsymbol{\xi}_2^{T} + \boldsymbol{\xi}_2\boldsymbol{\xi}_1^{T}),
\end{split}
\end{align}
where \(\boldsymbol{\eta}\) is the global axis ($\mathbf{e}_x$, $\mathbf{e}_y$, or $\mathbf{e}_z$ depending on the Co-Co bond) that is perpendicular to the Co--Co bond direction, and \(\boldsymbol{\xi}_1\), \(\boldsymbol{\xi}_2\) are the two remaining global axes. For next neighbor bonds, the isotropic contribution $\hat{J}_{ij} = J_2 \mathbb{I}$ is used.

Then, the minimization of Fourier transform of exchange matrix performed on a \(101 \times 101 \times 101\) grid in the Brillouin zone shows that the ground state of Sr$_2$CoTeO$_6$ corresponds to the propagation vector \(\mathbf{q}_{\min} = (0,0,-1/2)\), which confirms the type-I of antiferromagnetic ordering. It is worth noting that while present calculations identify importance of Kitaev interaction, further studies including computations of full exchange tensors are needed to explain experimentally observed structure with all spins canted on 58$^{\circ}$.

%%%%%%%%%%%%%%%%%%%%%%%%
\section{Altermagnetism}
\begin{figure}[htb]
\centerline{\includegraphics[width=\linewidth]{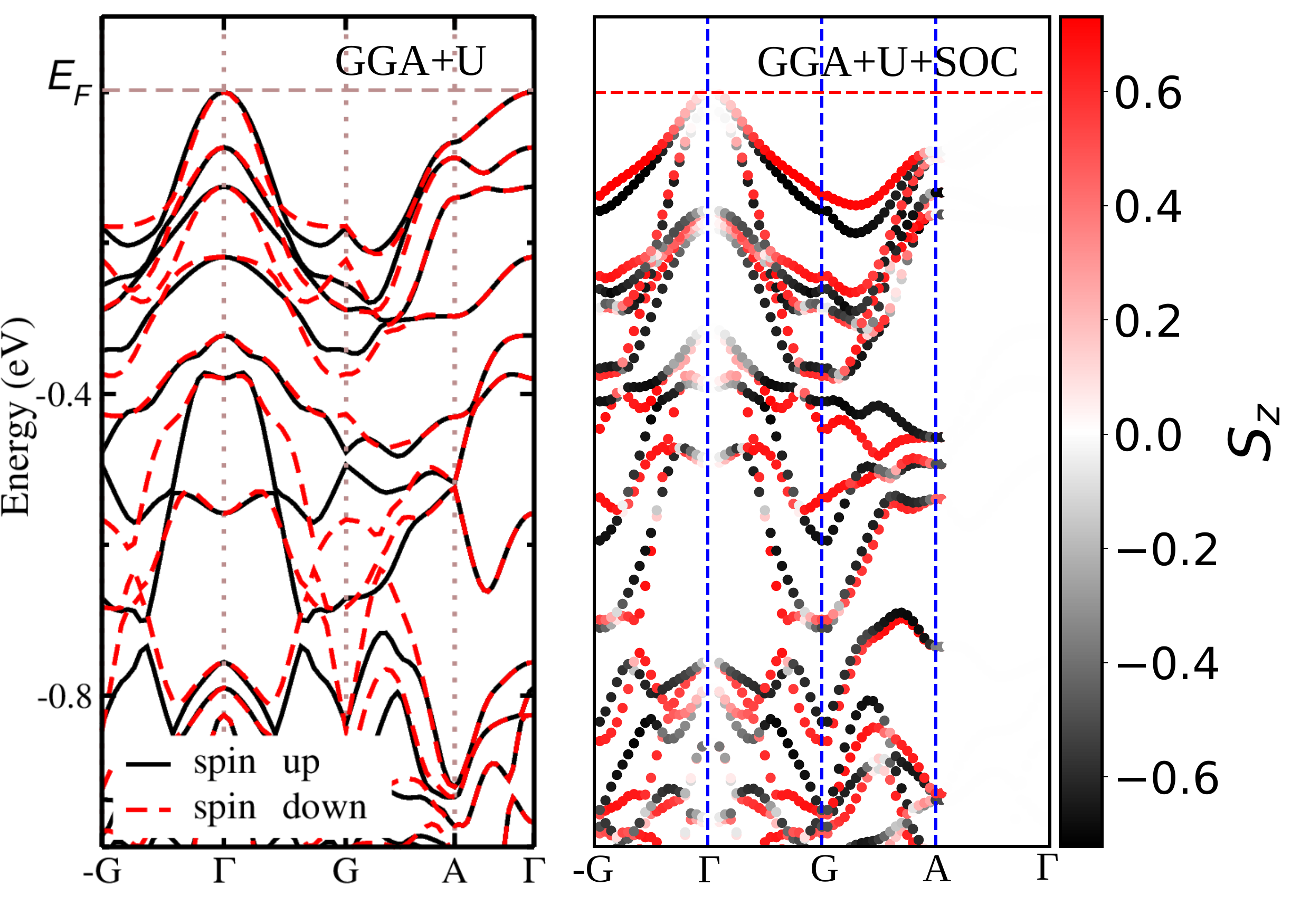}}
\caption{The band structure of Sr$_2$CoTeO$_6$ obtained in GGA+U and GGA+U+SOC ($S_z$ component) approximations in the energy range of [-1.5, 0.1] eV. The Fermi energy is in zero. Coordinates of high-symmetry $k$-points: $\Gamma=(0,0,0)$ and $\textrm{A}=(0,0,0.5)$, given in fractional coordinates of the reciprocal lattice. A generic point $\textrm{G}=(0.417, 0.237, 0.371)$ is also shown.\label{fig:bands}}
\end{figure}

In spite of the rather strong spin canting, the magnetic structure of Sr$_2$CoTeO$_6$ remains collinear. The magnetic point group is $2/m.1$~\cite{orayech2015}. The $PT$ symmetry is broken in this case, and one can expect altermagnetism~\cite{guo2023}. According to the classification by Cheong et al.~\cite{cheong2025} it must be the $M$-type, while the presence of the $2_{010}$ rotation axis in this group suggests $d$-wave altermagnetism in terms of the classification by \v{S}m\v{e}jkal et al.~\cite{smejkal2022a}.

Spin splitting in electronic bands persist both with and without spin-orbit coupling, thus, the system exhibits strong altermagnetism according to Ref.~\cite{cheong2025}. 
%It is more pronounced for $d$ states. In Fig.~\ref{fig:bands} the energy region, where occupied Co $3d$ states dominate.  
Indeed, one can see from Fig.~\ref{fig:bands} that while the spin-orbit coupling modifies the band structure, the spin splitting generally has a non-relativistic origin and is mostly defined by the exchange interaction.

The $2/m.1$ magnetic point group contains the following symmetry elements in addition to the unity transformation: $2_{010}$, $m_{010}$, and  $\bar 1$. The inversion $\bar 1$ does not flip pseudovectors such as spins, but maps $k \to - k$. This explains the spin splitting that is symmetric with respect to the $\Gamma$-point (e.g., along the generic path $\bar G - \Gamma - G$). The presence of the $2_{010}$ rotation axis about the $b$ axis together with $\bar 1$, forbids spin splitting along $k_z$ (i.e., along the $A-\Gamma$ direction).

Altermagnetism is known to induce various nontrivial physical properties, many of which have been, however, predicted many years by E.A. Turov, see e.g.~\cite{Turov-book} and rediscovered only recently. Based on symmetry considerations, one would expect the transverse piezomagnetic effect in Sr$_2$CoTeO$_6$ with the following symmetry allowed components of the piezomagnetic tensor $\Lambda$: $M_x=\Lambda_{14} \sigma_{yz}$, $M_y=\Lambda_{23} \sigma_{z}$, $M_y=\Lambda_{25} \sigma_{xz}$, and $M_z=\Lambda_{36} \sigma_{xy}$. Here, $M_i$ is the magnetization and $\sigma_{ij}$ is the stress. B.A. Tavger, V.M. Zaitsev, and  I.E. Dzialoshinskii pointed out on possibility piezomagnetism using symmetry arguments,\cite{tavger1956magnetic,dzialoshinskii1958problem} while  experimentally it was discovered by A.S. Borovik-Romanov\cite{borovik1960piezomagnetism}. The microscopic mechanism underlying the piezomagnetic effect in insulators is often related to the Dzyaloshinskii-Moriya interaction, where external stress lowers the crystal symmetry, thereby inducing a canting of the magnetic moments and a non-zero total magnetization. Since the effect of the spin-orbit coupling is known to be important for the Co$^{2+}$ ions, one might expect a sizable piezomagnetic response in this material.

\begin{figure}[htb]
\centerline{\includegraphics[width=\linewidth]{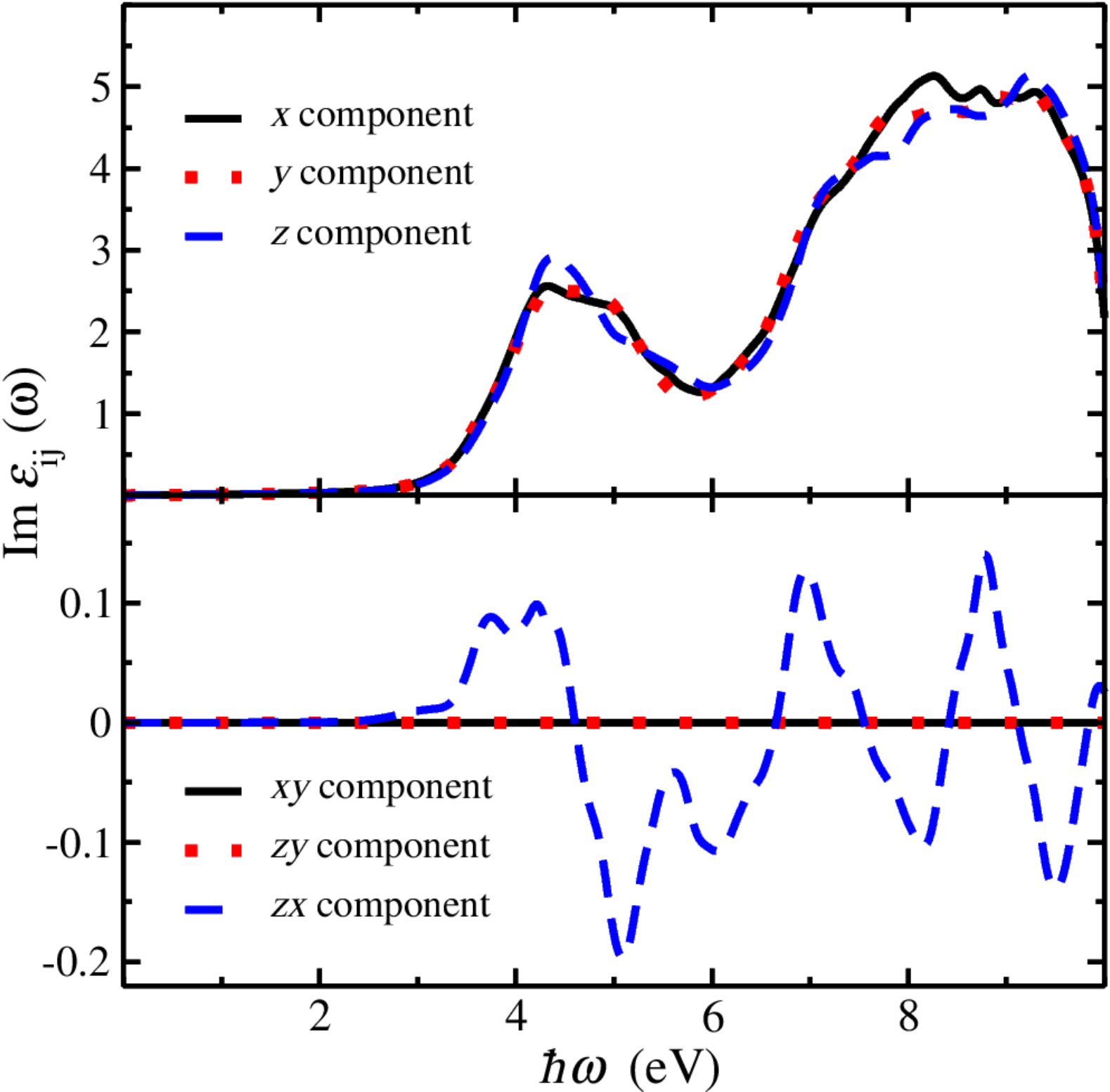}}
\caption{The imaginary part of different components of dielectric function $\varepsilon_\textrm{ij}$ obtained in GGA+U+SOC calculations.\label{fig:df}}
\end{figure}

%\section{Conclusions}\label{sec5}

Furthermore, a spontaneous optical Hall effect with a nonzero $\sigma_{xz}$ component of the electric conductivity would also be expected even in a gapped material, arising from interband transitions. In Fig.~\ref{fig:df} we show the imaginary part of the dielectric function $\textrm{Im}\,\varepsilon_{ij} (\omega)$, which is related to the real part of the frequency-dependent optical conductivity $\sigma_{ij}(\omega) \sim \omega\,\textrm{Im}\,\varepsilon_{ij}(\omega)$, where $\{i,j\} \in \{x,y,z\}$. Indeed, a finite $\sigma_{xz}(\omega)$ is clearly observed, while the other two off-diagonal terms vanish. Diagonal elements $\sigma_{ij}$ are allowed to be different by symmetry.

Therefore, it would be highly interesting to measure e.g. the Kerr effect in this compound. For a variation of the optical impermeability tensor $\Delta \beta_{ij} = q_{ijk} H_k$ with the magnetic field $H_k$, the symmetry-allowed nonzero components of $q_{ijk}$ are $q_{xxy}$, $q_{yyy}$, $q_{zzy}$, $q_{yzx}$, $q_{yzz}$, $q_{xzy}$, $q_{xyx}$, and $q_{xyz}$.
%%%%%%%%%%%%%%%%%%%%
\section{Conclusions}

In summary, our \textit{ab initio} study establishes Sr$_2$CoTeO$_6$ as a promising platform bridging Kitaev physics and altermagnetism. We find a substantial ferromagnetic Kitaev exchange for nearest neighbors, \(|K_1| \approx J_1/3\). Next nearest exchange interaction $J_2$ appears to be small and ferromagnetic, \(J_2/J_1 \approx -0.06\). The model including exchange parameters calculated by GGA+U+SOC approach reproduces experimentally observed magnetic order. Note, that these parameters also place the material close to the spin-liquid regime, which naturally explains its large frustration parameter (\(f \sim 10\)) and suppressed \(T_N\). This identifies a Co-based double perovskite as a new candidate Kitaev material beyond the traditional \(4d\) and \(5d\) compounds, and our results should be transferable to other Co-based double perovskites, including highly frustrated systems such as Ca$_2$CoTeO$_6$ and Sr$_2$CoNbO$_6$~\cite{Vasala2014a}. 

Furthermore, we demonstrate that Sr$_2$CoTeO$_6$ is a robust  altermagnet with spin splitting permitted by the \(2/m.1\) magnetic point group and confirmed by GGA+U+SOC calculations. Importantly, the coexistence of altermagnetism and sizable Kitaev exchange implies that applied stress not only induces a transverse piezomagnetic response but can also modify the ratio between the isotropic Heisenberg and Kitaev terms, since the two are governed by the same bond-dependent hopping pathways. This offers an additional handle for strain-engineering the magnetic ground state and proximity to the spin-liquid regime. The predicted non-vanishing off-diagonal optical conductivity \(\sigma_{xz}(\omega)\), with all other off-diagonal components forbidden by symmetry, further suggests a measurable Kerr effect, making Sr$_2$CoTeO$_6$ an interesting candidate for antiferromagnetic spintronics and magneto-optical applications.

\subsection*{Acknowledgments}
We are grateful to P.A. Maksimov for fruitful discussions.

\subsection*{Funding}
Analysis of magnetic properties of Sr$_2$CoSiO$_6$ was supported by the Russian Science Foundation via RSF 23-12-00159-P. Studying of altermagnetism was supported by Ministry of Science and Education of Russia via M.N. Mikheev Institute of Metal Physics.

\subsection*{Conflict of interest}
The authors declare no conflicts of interest.

\end{document}